# UBIQUITOUS SMARTPHONE BASED LOCALIZATION WITH DOOR CROSSING DETECTION

Jan Racko [a], Juraj Machaj [a,*], Peter Brida [a]

[a] Department of Multimedia and Information-Communication Technologies, Faculty of Electrical Engineering, University of Zilina, Univerzitna 1, 010 26 Zilina, Slovakia
{jan.racko, juraj.machaj, peter.brida}@fel.uniza.sk
*Corresponding author: juraj.machaj@fel.uniza.sk

***Abstract:*** *The positioning of users using their smartphones represents interesting service for various areas. Position of users can represent valuable information for various service providers. In industry 4.0 smart devices such as smartphone or tablet can help users to react faster in case of emergency. It can help to reduce reaction times for service workers and improve users' orientation in the environment, or delegate work to the nearest available employee. In previous work we have described modular positioning system that is able to automatically select optimal positioning module based on available radio signals. In this paper we will focus on estimation of crossing between environments e.g. indoor and outdoor. The proposed algorithm is crucial for implementation of seamless positioning. Since both environments have specific demands not only on positioning solutions but also on maps and navigation algorithms. The proposed algorithm does not require deployment of new infrastructure and can be used on widely available smartphones. The algorithm was tested in real conditions by different users and using various smartphone devices to prove its performance under various conditions.*

## 1. INTRODUCTION

Currently positioning of mobile devices represent important topic not only for researchers, and service providers, but also for industry (Rojko A., 2017). Nowadays every employee has a smart device that could help him to improve his work performance by introducing new location based services (LBS). In industry 4.0 there is a possibility to utilize LBS in order to reduce time required to fix broken equipment, delegate work to the nearest employee, improve orientation of (new) employees in the environment, or in case of emergency situations. Such services will, however, require robust positioning system that will be able to work seamlessly in both indoor outdoor environments.

Currently there is a large number of specialized systems developed for indoor environment (Xiao J., et al., 2016, Yassin A., et al., 2017). These are based mainly on signals from available wireless networks (Cipov, V., et al. 2012, Kriz P., et al., 2016) and data from inertial measurement units (IMUs) (Yassin A., et al., 2017, Bojja, J., et al. 2016), although other types of data like visible light (Qiu, K., et al. 2016), ultrasound (De Angelis G., et al., 2015) or camera images (Jiao J., et al., 2017) can be utilized to estimate position of mobile users and devices.

On the other hand, in the outdoor environment we have well established satellite positioning systems, commonly called Global Navigation Satellite Systems (GNSS). GNSS are in theory able estimate position of user anywhere in the world. They perform reasonably well, when signals from satellites are not blocked by obstacles or affected by multipath propagation phenomenon. Unfortunately, these conditions are hard to achieve in urban areas or industrial parks, therefore, in some cases alternative positioning systems have to be used also in the outdoor environment (Kriz P., et al., 2016).

In the previous work we have proposed modular positioning system (Brida, P., et al. 2014), which can be used in both indoor and outdoor environments with a good positioning accuracy. The idea of the modular positioning system is to estimate positions of users by automatically selected positioning solution based on available radio signals and data from available sensors. Currently the system allows positioning using Wi-Fi, cellular networks and GNSS (Brida, P., et al. 2014). Moreover, IMUs integrated in the modern smart devices can be utilized to improve positioning accuracy for dynamic users.

In this paper we will focus on problem of detection of crossing between environments, e.g. users moving from indoor to outdoor environment and vice versa. The algorithm for door detection will be proposed and tested for this purpose. The proposed solution does not require any additional infrastructure and is based on sensors commonly implemented in smartphones. The algorithm will allow not only automatic switching between maps for indoor and outdoor environments but could also lead to further improvements of modular positioning system from the accuracy point of view, as more information about environment and movement of user will be available.

The rest of the paper is organized as follows; related work is presented in the Section 2. The Section 3 describes the proposed algorithm for door detection. Experiments and achieved results will be presented in the Section 4 and the Section 5 will conclude the paper.

## 2. RELATED WORK

### 2.1 Modular Positioning System

The modular positioning system is currently under development at the Department of Multimedia and Information-Communication Technologies (Brida, P., et al. 2014). The main idea is to develop single localization system that will be able to provide position estimates seamlessly in heterogeneous environments e.g. indoor and outdoor. It is based on the assumption that in various environments various approaches to position estimation have to be used. The system is able to provide user with position estimates in both indoor and outdoor environments, using GPS, Wi-Fi or GSM signals, as can be seen from Figure 1. The system is able to automatically select optimal positioning solution based on measured signals taking into account number of signals and their quality. Moreover, IMU can be used to estimate path of the user. The modules were chosen based on the fact that these signals are available at all devices. Since the modular system is an open platform other modules can be implemented in the future, based on available technologies, e.g Bluetooth, Zig-Bee, UWB, etc.

**Figure 1.** Modular Positioning System

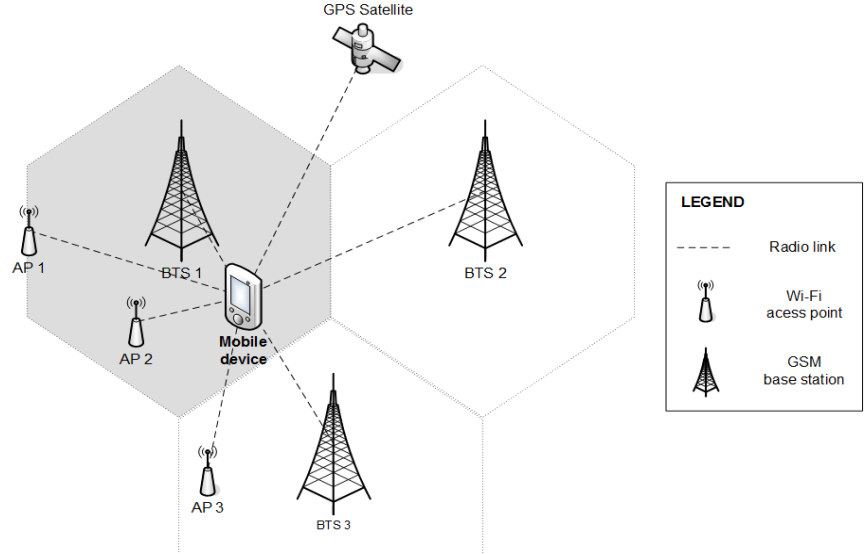

Positioning modules that use signals from Wi-Fi and GSM networks are based on fingerprinting approach. The fingerprinting framework is based on assumption that available signals and their parameters depend on the position of user. The framework requires so called calibration phase (or off-line phase). During the calibration phase the measurements of Received Signal Strength (RSS) are collected and stored in the radio map database. The radio map database consists of reference points with known position linked to RSS samples measured at the given reference point.

During the positioning phase (or on-line phase) the position of mobile user is estimated based on comparison between on-line measurements with the data stored in radio map database. Euclidean distance is most widely used for comparison of RSS samples (Machaj J. and Brida P., 2011) and position estimator can be defined as:

$$\hat{x} = \frac{\sum_{i=1}^{M} \omega_i \cdot c_i}{\sum_{i=1}^{M} \omega_i}, \qquad (1)$$

where $c_i$ represent position of $i$-th reference point $M$ is number of reference points and $\omega_i$ is a non-negative weighting factor. Traditionally the weights are computed as inverted value of distance between RSS vectors. The estimator of the formula (1), which keeps the K highest weights and sets the others to zero is called the WKNN (Weighted K-Nearest Neighbours) method (Bahl P. and Padmanabhan V. N., 2000). WKNN with all weights $\omega_i = 1$ is called the KNN (K-Nearest Neighbours) method and the simplest method, where $K = 1$, is called the NN (Nearest Neighbour) method (Tsung-Nan L. and Po-Chiang L., 2005). In (Honkavirta V., et al. 2009) it was found that WKNN and KNN methods commonly outperform the NN method, especially in cases when $K$ is set to 3 or 4.

### 2.2 Door detection algorithms

Algorithms for door detection are becoming important part of positioning and navigation systems in indoor environment, since their use can further improve accuracy of existing systems. In this subsection a short overview of door detection algorithms will be presented.

The door detection algorithms can be in principle divided into three main categories, the first one are vision based solutions. These are mainly based on image from camera (Adar U. G. and Bayindir, L., 2015, Shalaby M. M., et al. 2014, Kim, S., et al. 2011) and developed for application in robotics, since robots are commonly equipped with at least one camera. More recently approaches based on laser scanning (Quintana B., 2016) and depth measurements using a Kinect camera (Dai D., et al. 2013) have also emerged. Camera based door detection is mostly based on approaches that utilize corner and shape detection algorithms in combination with image classification. These approaches can provide good results, however, are not optimal for use with smartphone camera due to the fact that most users hold the smartphone in a way that its back and front cameras are facing down and up, respectively. Therefore, captured images lack data required for door detection.

The second category of door detection algorithms are infrastructure based solutions, these algorithms utilize data from new infrastructure that has to be deployed on site. This means that implementation cost might rise significantly and, on top of that, widely used smart devices might not be equipped by sensors or receivers that are required for system to operate. Such solutions are most widely based on RFID tags (Xie, X., et al. 2015) or UWB transmitters (García, E., et al. 2015). As these technologies might provide high positioning accuracy.

The third category of algorithms is represented by infrastructure free approaches. Algorithms from this category mainly use data that are provided by sensors integrated in smart devices and therefore does not require further investments into new infrastructure or new devices. Most of the proposed solutions presented in available literature was based on magnetometer readings.

Door detection algorithm available on smartphone devices was presented in (Zhao, Y., et al. 2015), authors used raw magnetometer readings to estimate crossing of the doors. The main advantage of the approach is that the algorithm does not need any information about floor plan or any calibration measurements. The approach can provide detection of the doors with success rates around 60% and depends highly on door material.

MagicFinger (Carrillo, D., et al. 2015) is a representative solution of magnetic based positioning system. The approach is based on 3D magnetic fingerprinting and can provide reasonable positioning accuracy. However, using magnetic fingerprinting for door detection requires database of magnetic fingerprints of all doors measured in different directions as magnetic field depends on orientation (Torres-Sospedra J., et al., 2015). Moreover, comparison of magnetometer output data with the database is not trivial since magnetic signatures are not measured on static points but should be measured during movement, therefore difference of movement speeds between data in magnetic map and data measured during positioning has to be considered.

## 3. PROPOSED ALGORITHM

The proposed algorithm is based on Pedestrian Dead Reckoning (PDR) which utilizes smartphone IMU sensors to estimate path of the user. The proposed algorithm for door crossing detection is using tracking information together with data from IMU sensor to estimate if user crossed door or not. In the paper we focus primarily on non-automatic door.

### 3.1 Pedestrian Dead Reckoning

In the proposed algorithm we are using approach where heading angle is calculated from gyroscope data, and step detection is done by using output data from accelerometer. Particle filter (PF) has been used in the indoor environment as a correction method. More details on calculation process can be found in our previous work (Racko, J., et al. 2016). As described in our previous work, we have used fixed step length of 0.75 m as this value was proven to provide reasonable results with combination of PF. Incorrectly chosen step length may lead to increased positioning error. Use of PF is feasible in the indoor environment, because there were many obstacles which helps to eliminate some particles used for positon estimation, and thus improve position estimation accuracy.

On the other hand use of magnetometer for corrections could lead to higher errors in path estimation because of metal parts in building, such as railings, or doorframes. On the contrary, in the outdoor environment use of magnetometer data for estimated path correction seems to be a better option. PF will not be able to provide good correction since there are no fixed obstacles, which could eliminate particles used to estimate position of the user.

Moreover, we have used Kalman filter (KF) to fuse gyroscope and magnetometer data. Principle of Kalman filter fusion can be found in (Abyarjoo F., et al. 2015). Normalized acceleration is calculated by using:

$$a_{norm}(t) = \sqrt{a_x^2(t) + a_y^2(t) + a_z^2(t)} - g, \qquad (2)$$

where $a_{norm}$ represents normalized acceleration, $a_x$, $a_y$, $a_z$ denotes accelerations in all three axes respectively in time $t$ and $g$ represents gravity acceleration. In Figure 2 example of normalized acceleration is shown over time, these data are part of measurements performed during testing. From the figure it can be noticed that at some area level of acceleration drop, the area is marked by red square.

**Figure 2.** Level drop of normalized acceleration due to door opening

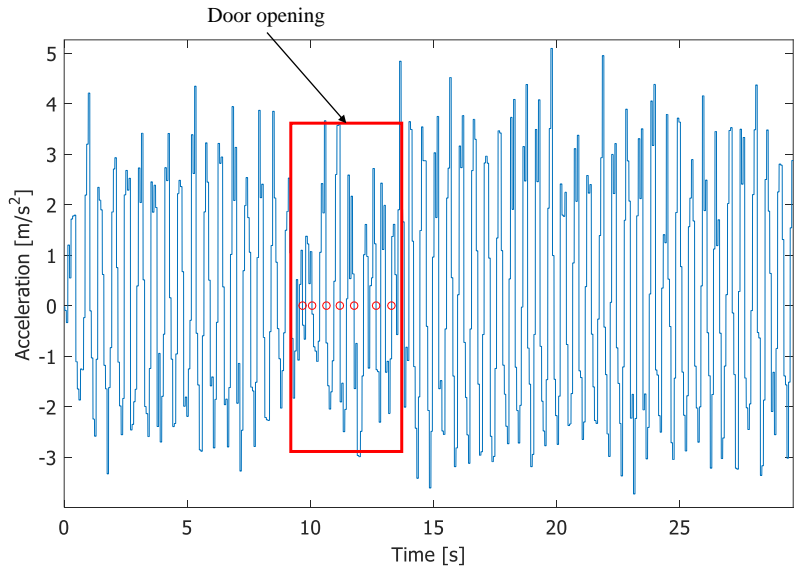

The marked area represents phase of door opening. Before door opening pedestrian has to slow down or even stop, however, some minor moves are still detected. Thanks to this we assume that it is possible to detect when pedestrian reaches doors. To detect door opening upper and lower threshold for acceleration have been set to 0.5 and -0.5 m/s². If acceleration exceeds these maximum and minimum thresholds, but doesn't reach levels for step detection, the algorithm assumes that user has reached doors. Step detection thresholds were set to 1.5 and -1.5 m/s². In Figure 2 red circles represent zero crossings that occurred during the door opening.

## 3.2 Cross door detection

To evaluate whether user crossed doors it was needed to define crossing areas and crossing zones in indoor environment and outdoor environment, respectively. The difference between crossing area and crossing zone can be seen in Figure 3.

**Figure 3.** Difference between crossing area and crossing zone

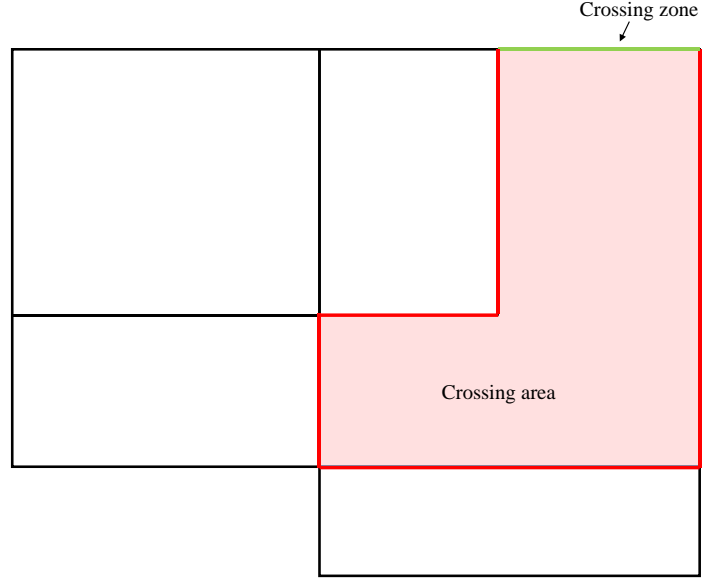

The basic assumption for door crossing estimation is that information about door position is available. Common width of doors is 90 cm, however, from our experiment, which will be shown in next section, it is clear that this width is not sufficient for detection due to positioning errors, therefore size of 5 m has been chosen to define crossing zone. The zone was defined by 2.5 m distance to both directions from door´s center point.

In the indoor environment position estimation has been done using PDR corrected by PF. Although, in the outdoor environment PDR was used and heading angle was corrected by fusion of gyroscope data with magnetometer data by KF.

If the system detects that pedestrian reached crossing area, door detection algorithm will be started automatically. The algorithm also automatically identifies crossing point between estimated path and crossing zone. This is done by calculation of intersection point between two lines. First line L1 represents crossing zone and second line L2 represents path segment between actual and previous step positions, the crossing point then can be calculated from:

$$\begin{aligned} x &= \frac{(x_1 y_2 - y_1 x_2)(x_3 - x_4) - (x_1 - x_2)(x_3 y_4 - y_3 x_4)}{(x_1 - x_2)(y_3 - y_4) - (y_1 - y_2)(x_3 - x_4)} \\ y &= \frac{(x_1 y_2 - y_1 x_2)(y_3 - y_4) - (y_1 - y_2)(x_3 y_4 - y_3 x_4)}{(x_1 - x_2)(y_3 - y_4) - (y_1 - y_2)(x_3 - x_4)} \end{aligned}, \quad (3)$$

where $[x, y]$ are coordinates of crossing point, $[x_1, y_1]$, $[x_2, y_2]$ are coordinates of crossing zone line L1 and $[x_3, y_3]$, $[x_4, y_4]$ are coordinates of line L2 which is defined by actual and previous step.

Only in case that both crossing point and doors were detected and both detections have been done within 5 steps, switching between environments will be initialized. If these conditions are not met, algorithm will not initialize switching between environments.

## 4. EXPERIMENTAL SETUP

The experiments were focused on effectivity of the proposed algorithm. All experiments have been done by holding a smartphone in hand. Our first measurements were focused to set crossing zone width. We have chosen two points close to the entrances of buildings, and track between them was walked 50 times from the starting point to the ending point with three different smartphones. Used devices were Samsung Galaxy S6 Edge (SGS6E), Samsung Galaxy Note 4 1(SGN4) and XiaoMi Max (XMM). The starting point marked with red circle and the ending point marked with green circle can be seen In the Figure 4. The blue line in the figure denotes experimental path.

**Figure 4.** Reference path with starting and ending point

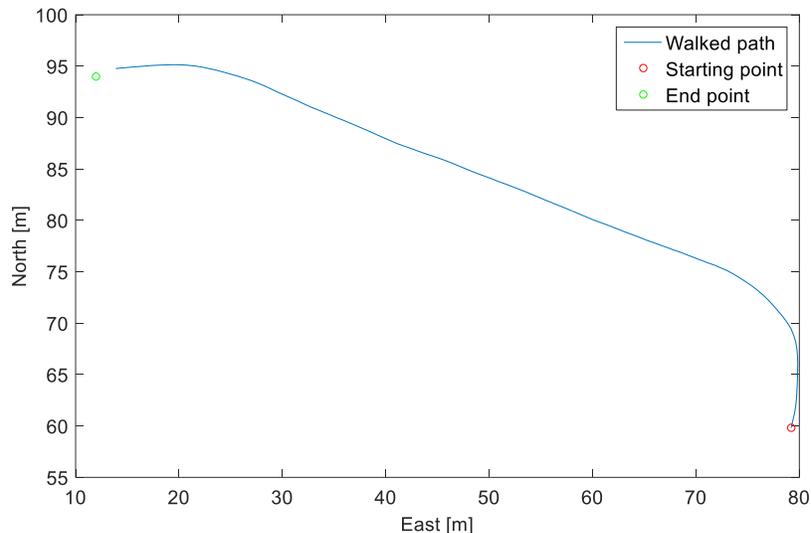

For testing of the proposed algorithm two entrances into buildings of the University of Žilina have been chosen. The first entrance was in building of the Faculty of Electrical Engineering and second was into building of the Faculty of Operation and Economics of Transport and Communications. Position of both entrances was measured by precise GPS receiver and WGS84 coordinates were transformed to Cartesian coordinates. In Figure 5 the testing path is shown where blue line represents chosen path and red points denotes crossing zones.

**Figure 5.** Walked path with crossing zones

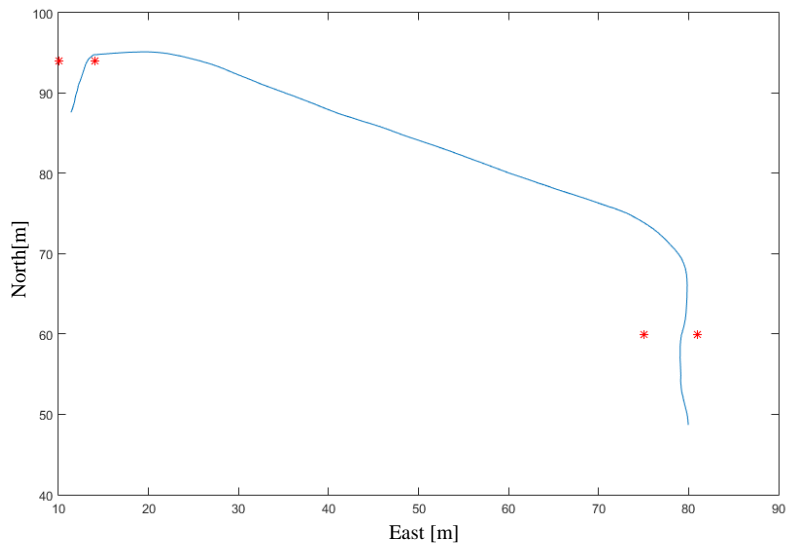

Firstly the test path was walked with mentioned smartphones held in hand by one reference person. Then same path was then walked by three different persons, with the reference smartphone SGS6E. For all cases, chosen path was walked 50 times. This was decided in order to have sufficient amount of data to evaluate performance of door crossing algorithm. Since all measurements combined sum up to 300 trials and there are 2 door crossings in each trial, giving us total of 600 door crossings (100 door crossings per case). The path started in the indoor environment at the Faculty of Electrical Engineering, and ended again indoors at the Faculty of Operation and Economics of Transport and Communications.

Third experiment was aimed on case when pedestrian goes from the indoor environment in direction towards the door, however, turns back just in front of doors, as can be seen in Figure 6. This experiment was done only with one smartphone SGS6E, and with one person. The test was aimed to evaluate reliability of proposed algorithm.

**Figure 6.** Path during the third experiment

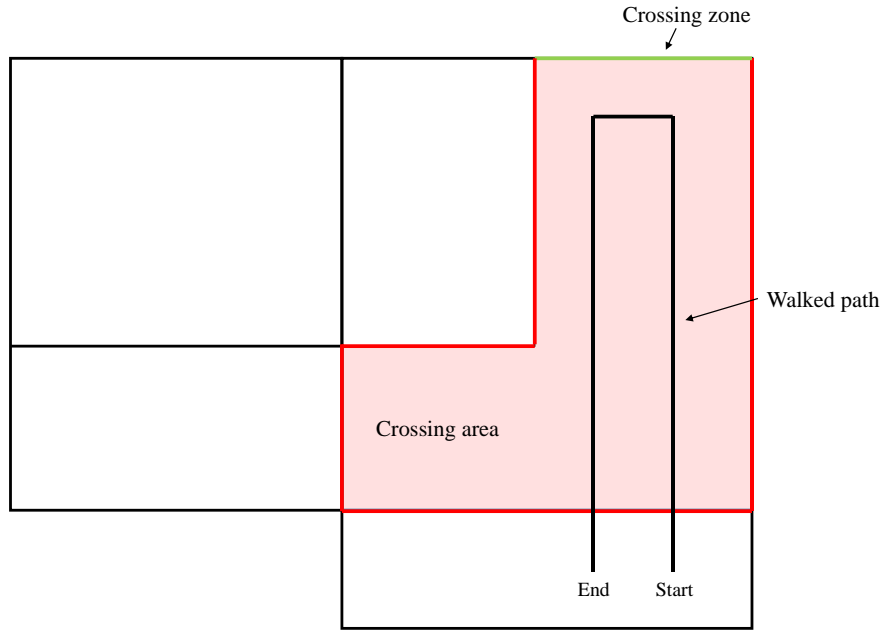

## 5. ACHIEVED RESULTS AND DISCUSSION

The main goal of the first experiment was to estimate positioning error in order to set the size of the crossing area. Statistical parameters of the achieved positioning error from the first set of measurements can be seen in the Table 1. The errors presented in the table are calculated from 50 independent trials for each smartphone and represent localization error at final position, i.e. difference between final point and final estimated position. Since error of the PDR algorithm is integrated over the time, these values represent the highest error achieved during individual trials.

Table 1: Localization error

| **Localization error [m]** | | | |
|---|---|---|---|
| **Device** | **min** | **max** | **average** |
| **SGS6E** | 0.02 | 12.05 | 2.6 |
| **SGN4** | 0.1 | 18.96 | 3.5 |
| **XMM** | 0.25 | 6.01 | 2.78 |

For better representation CDF functions of positioning errors for each smartphone are shown in Figure 7. From the CDF functions it is obvious that 70 % of errors were lower than 4 m, and for 80 % of results was error less than 5 m. Based on the test results we decided to set the width of crossing zones to 5 m.

**Figure 7.** CDF of achieved localization errors for individual devices

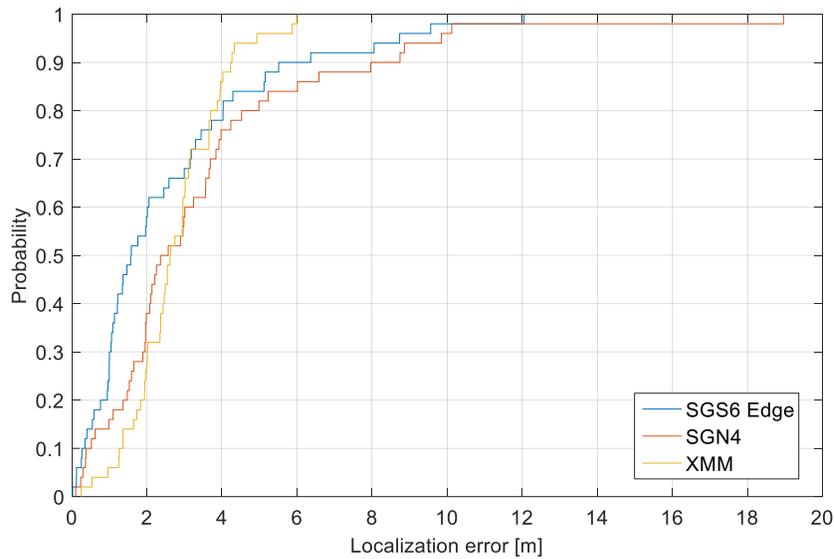

During the following tests the proposed algorithm was tested on both using different devices and the same device by different people. Firstly the path shown in the Figure 5 was walked with one person and three different smartphones. For each walk there were only two crossings. From the Faculty of Electrical Engineering to outdoor environment and the second was from the outdoor to the Faculty of Operation and Economics of Transport and Communications. Normalized acceleration can be seen in the Figure 8. From the figure it is obvious that there are two acceleration level drops which represent crossing of doors. In upper line with red circles are marked detected doors and in lower line with purple circles are shown detected door crossing points.

**Figure 8.** Normalized acceleration with door and door crossing detection

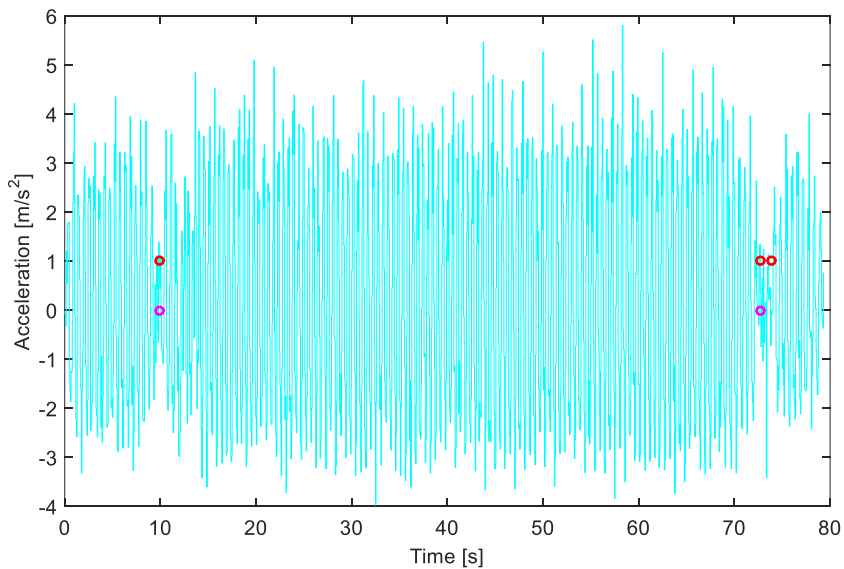

To verify robustness of the algorithm same path was walked with different pedestrians using the reference device. Achieved results showing effectivity of switching between environments are shown in the Table 2.

Table 2: Results achieved by the proposed algorithm

| Effectivity of switching [%] | | |
|---|---|---|
| **Pedestrian 1** | **Pedestrian 2** | **Pedestrian 3** |
| 87.6 | 77.2 | 83.4 |
| **Device 1** | **Device 2** | **Device 3** |
| 85.1 | 84.3 | 78.6 |

Combination of door detection from accelerometer signal combined with cross door detection represents reliable approach, with success rate close to 80 %. Last experiment was when pedestrian turned back in front of doors, this experiment was to estimate robustness of the algorithm for classification of negative events, i.e. walking in the crossings area without crossing the door. Starting position was in the indoor environment and pedestrian turned back before he reached the door. This represent very common case when pedestrian has to turn back because he forgot something in the building. Described experiment was tested 50 times with one smartphone SGS6E and with one pedestrian and we reach effectiveness of 98 %. Even though PDR provided position estimates that would place pedestrian to the outdoor environment. However, pedestrian just turned back, therefore no doors have been detected so algorithm did not initiated switching between environments. In this experiment the switching was occurred only in case when pedestrian had to slow down and avoid other persons walking in or outside the building. All the classification results achieved by door detection algorithm are summed up in the Table 3.

Table 3: Confusion matrix for door crossing detection based on all performed tests.

| | | Actual value | |
|---|---|---|---|
| | | **Positive** | **Negative** |
| **Estimated value** | **Positive** | 82.7 % | 2 % |
| | **Negative** | 17.3 % | 98 % |

From the table it can be seen that proposed algorithm was able to detect door crossing in 82.7 % of time, however, it was able to correctly recognize that person did not get thru the door in 98 % of cases. The lower percentage of true positive detections is caused by localization error, caused by noisy data from IMU sensors. The error is in this case integrated over whole track, since position of user is estimated only using PDR. With regular update of the position from other positioning system the error should be reduced and classification of positive door crossings might be increased as well.

## 6. CONCLUSION

In the paper novel algorithm for detection of crossing between environments was proposed. The main idea behind the proposed algorithm is to utilize information from PDR together with normalized acceleration to detect door opening by moving pedestrian. The algorithm is automatically triggered when user come to the crossing area and check if he crossed thru the crossing zone. The crossing zone represents area of doors and was set to 5m based on experimentally achieved localization error.

The proposed algorithm was tested in three experiments. In the first two experiments effectivity of algorithm was tested by one user and different devices and different users and one device in order to see what the impact of both user and device have on performance. From the results it seems that both device and user have approximately the same impact on the achieved accuracy. On average success rate of crossing between environments detection was above 82 %, false negative error was approximately 17 %.

In the last experiment robustness of the proposed algorithm has been tested. User came to the crossing area but turned back just before the doors. In this case, the algorithm achieved success rate of 98 % and the false positive error was only 2 %. From presented result it is clear that our algorithm can be used as suitable solution for switching between different environments.

## ACKNOWLEDGEMENT

This work has been partially supported by the Slovak VEGA grant agency, Project No. 1/0263/16 "Research of integrated positioning system based on wireless systems and sensors implemented in intelligent mobile devices" and European Union's Horizon 2020 research and innovation programme under the Marie Skłodowska-Curie grant agreement No 734331 and by project ITMS: 26210120021, co-funded from EU sources and European Regional Development Fund.